# Image Steganography: Protection of Digital Properties against Eavesdropping

Ajay Kumar Shrestha
Computer and Electronics Engineering
Kantipur Engineering College, TU
Lalitpur, Nepal
ajayshrestha1@gmail.com

Ramita Maharjan
Computer and Electronics Engineering
Kantipur Engineering College, TU
Lalitpur, Nepal
ramitamaharjan@kec.edu.np

Rejina Basnet
Computer and Electronics Engineering
Kantipur Engineering College, TU
Lalitpur, Nepal
rejinabasnet@kec.edu.np

*Abstract*—Steganography is the art of hiding the fact that communication is taking place, by hiding information in other information. Different types of carrier file formats can be used, but digital images are the most popular ones because of their frequency on the internet. For hiding secret information in images, there exists a large variety of steganography techniques. Some are more complex than others and all of them have respective strong and weak points. Many applications may require absolute invisibility of the secret information. This paper intends to give an overview of image steganography, it's uses and techniques, basically, to store the confidential information within images such as details of working strategy, secret missions, criminal and confidential information in various organizations that work for the national security such as army, police, FBI, secret service etc. The desktop application has been developed that incorporates Advanced Encryption Standard for encryption of the original message, and Spatially Desynchronized Steganography Algorithm for hiding the text file inside the image.

## I. INTRODUCTION

Since the rise of the internet, the security of information has also been one of the most challenging factors of information technology and communication. Huge volume of data is transferred every second in the internet via e-mails, file sharing sites, social networking sites etc. As the number of internet users rises, the concern on the credibility of the services is also on rise, so the concept of internet security has become the important research topic nowadays. The competitive nature of the computer industry has forced the web services into the market at a breakneck pace giving a very little time for audit of system security. On the other hand, the tight labor market causes internet project development to be managed with less experienced personnel, who may have no training in privacy and security issues. This combination of market pressure and the low unemployment creates an environment with machines vulnerable to exploits, and malicious users to intrude those machines. Due to the rapid development of communication technologies, it is convenient to acquire multimedia data. However, the problem of illegal data access occurs unexpectedly every time and everywhere. Hence, it is very important to protect the digital content while ensuring the users' privacy and the authorized use of multimedia data.

Data encryption is basically a strategy to make the data unreadable, invisible or incomprehensible during transmission by scrambling the content of data. Steganography is the art and science of hiding information by embedding messages within seemingly harmless messages. It also refers to the "Invisible" communication. The power of Steganography is in hiding the secret message by obscurity, hiding its existence in a non-secret file. Steganography works by replacing bits of useless or unused data in regular computer files. This hidden information can be plaintext or ciphertext and even images. According to Bhattacharyya et al., "Steganography's niche in security is to supplement cryptography, not replace it. If a hidden message is encrypted, it must also be decrypted if discovered, which provides another layer of protection" [1]. Generally, an Image Steganography uses some reliable algorithms or secret keys to transform or encrypt secret images into ciphered images. Only the authorized users can decrypt secret images from the ciphered images. The ciphered images cannot be recognized by unauthorized users who grab them without knowing the decryption algorithms. This prevents information leaks about their ongoing investigation on cases, task assignment, working strategy, confidential database etc.

The available desktop applications for the organizations working in crime and security sectors which have been following various techniques to communicate like emails, telephones etc., are more prone to attacks by the intruders including eavesdroppers. The eavesdropping is basically an attack on the transmitted data in order to capture the packets and retrieve the data content such as passwords, session tokens, or any kind of sensitive information. Tools like network sniffers are used to initiate such attacks by collecting the packets on the network and then analyzing the collected data for protocol decoding or stream reassembling.  So the confidential message to be transferred is often encrypted so that even if the outsiders succeed to receive the message, they become unable to read the original content. However, they may still know that the ciphertext contains some confidential information. Now, the concern has been driven towards steganography whereby one can hide secret information inside other cover objects which can be anything in today's digital world such as image, audio, video etc. Keeping this in mind, we have done this research work in order to hide confidential messages forwarded from one place to another by encrypting and then hiding them inside an image file so that no information leaks occur. Despite the fact that cryptography and steganography are not the same, we applied the combination of both the techniques in order to achieve better results for security. In our research work, we have also developed a desktop application for the secret organizations like police service where the communication needs to be confidential. The application provides a user friendly environment for the secret service personnel to type their messages, encrypt the written text and hide them inside an image file. The receiver can on the other side can also use the application to retrieve the .txt file hidden inside the image and hence decrypt the actual message from the retrieved file containing encrypted message. There are different encryption algorithms and the steganography techniques available so far. This research work has evaluated their performances and used Advanced Encryption Standard (AES) for encryption of the original message, and Spatially Desynchronized Steganography Algorithm (SDSA) for hiding the .txt file inside the image.

II. OVERVIEW

## A. Cryptography

The process of encryption involves disguising a message or making it cryptic so as to hide its actual appearance and meaning. Confidentiality, authentication, integration and non-repudiation are the characteristics of secured communication and cryptography guarantees the achievement of these. The general concept of cryptography is that for a given plaintext or message "M", we have an encryption function or algorithm "E" and key "K" to produce ciphertext or encrypted message "C" which can be decrypted using decryption function "D" to get the plaintext back. Basically there are two types of algorithms used in cryptography which are: Symmetric Algorithms and Asymmetric Algorithms.

*i. Symmetric Algorithms (Secret-key Algorithms)*

In symmetric algorithms, also known as conventional algorithms, the encryption key can be calculated from the decryption key and vice versa. Most symmetric algorithms use the same, secret key for both encryption as well as decryption. For this reason, these algorithms are popularly known as secret-key algorithms, single key algorithms, or one-key algorithms. The security of a symmetric algorithm entirely resides in the key; divulging the key means that messages are easily available for intruders. As long as the communication needs to remain secret, the key must remain secret. Encryption and decryption with a symmetric algorithm are denoted by:

$E_K(M) = C$

$D_K(C) = M$

Symmetric algorithms can be divided into two categories: stream ciphers and block ciphers. Stream ciphers operate on the plaintext, a single bit at a time, whereas the other operates on the plaintext in groups of bits also called blocks. For modern computer algorithms, a typical block size of 64 bits is large enough to preclude analysis and small enough to perform some work [2]. AES, DES, 3DES etc. are some important symmetric key algorithms available so far.

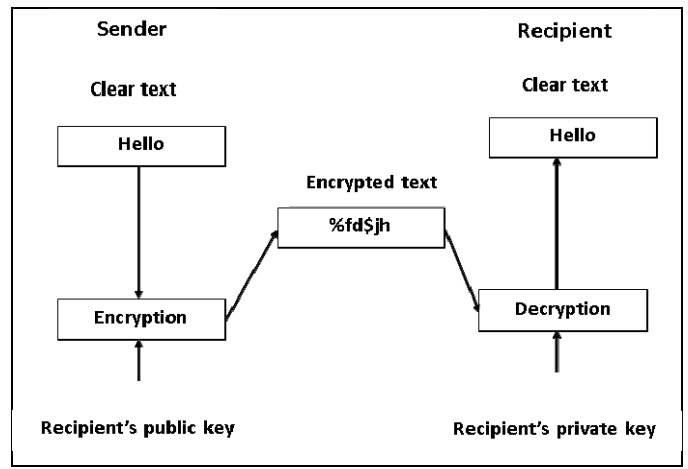

Fig. 1. Asymmetric Key Algorithm

*ii. Asymmetric Algorithms (Public-key Algorithms)*

Unlike, symmetric algorithms, the asymmetric-key algorithms use a key for encryption, which is different from the key used for decryption. Furthermore, the decryption key cannot (at least in any reasonable amount of time) be calculated from the encryption key and is kept secret. However, the encryption key can be made public as described by the name, "public-key algorithm". In other words, a complete stranger can utilize public encryption key to encrypt the message, but only a specific person who owns the corresponding decryption key can decrypt the message. Fig. 1 clearly demonstrates the asymmetric key algorithm.

In these systems, the encryption key is often called the public key "$K^+$", and the decryption key is often called the private key or secret key "$K^-$". Encryption using public key $K^+$ and decryption using private key $K^-$ is denoted by:

$E_{K+}(M) = C$

$D_{K-}(C) = M$

Contrary to this, digital signatures use private key for encryption and public key for decryption of messages. RSA is an example of public key algorithm.

## B. Cryptanalysis

The process of recovering the original plaintext from the encrypted message without access to the key is known as cryptanalysis. Successful cryptanalysis may recover the plaintext itself or the key used in the cryptography process. The loss of a key through non-cryptanalytic methods is called a compromise. An attempted cryptanalysis is called an attack. If others can't break an algorithm, even with knowledge of how it works, then they certainly won't be able to break it without that knowledge [2].

Generally, four types of cryptanalytic attacks are prevalent and each of them is based on the assumption that the cryptanalyst has the complete knowledge of the encryption algorithm.

*i. Ciphertext-only attack*

As the name suggests, in this type of attack, the cryptanalyst has only the ciphertexts of several messages, all of which have been encrypted using the same encryption algorithm. The ultimate goal is to recover the plaintext of as many messages as possible, or better to discover the key (or keys) used to encrypt the messages so that it can be used further to decrypt other messages encrypted with the same keys.

Given: $C1 = E_K(P1)$, $C2 = E_K(P2)$,...$Ci=E_K(Pi)$
Deduce: Either $P1, P2,...Pi$; $K$; or an algorithm to infer $Pi+1$ from $Ci+1= E_K(Pi+1)$

*ii. Known-plaintext attack*

In addition to the ciphertext, the plaintext of those messages are also accessible for the cryptanalysis process. The job is to deduce the key (or keys) used to encrypt the messages or an algorithm, in order to decrypt any new messages encrypted with the same key (or keys).

Given: $P1, C1 = E_K(P1)$, $P2, C2 = E_K(P2)$,...$Pi, Ci=E_K(Pi)$

Deduce: Either $K$, or an algorithm to infer $Pi+1$ from $Ci+1 = E_K(Pi+1)$.

*iii. Chosen-plaintext attack*

The cryptanalyst has access to both the ciphertext and associated plaintext for several messages and can choose the plaintext that gets encrypted. The analyst has to deduce the key (or keys) used to encrypt the messages or an algorithm, to decrypt any new messages encrypted with the same key (or keys).
Given: $P1$, $C1 = E_K(P1)$, $P2, C2 = E_K(P2)$,...$Pi, Ci = E_K(Pi)$, where the cryptanalyst gets to choose $P1, P2,...Pi$
Deduce: Either $K$, or an algorithm to infer $P i+1$ from $Ci+1 = E_K(Pi+1)$.

*iv. Adaptive-chosen-plaintext attack*

This is a special case of a chosen-plaintext attack. Apart from the features already present in chosen-plaintext attack, this one adds up little more. The cryptanalyst is allowed to choose a plaintext and also modify the choice based on the results of previous encryption. The analyst can choose one large block of plaintext to be encrypted and then a smaller block of the plaintext and then choose another based on the results of the first, and so forth making attack more efficient in search of the key or the algorithm itself.

C. *Spatial Domain Steganography*

Spatial domain steganography conceals the message bits in the images using the gray levels and the color values of the pixels directly. In terms of embedding and extraction complexity, these techniques are some of the simplest schemes. Even after the encoding of the message into image pixels, the change in image is not detectable to human eyes. The amount of additive noise that creeps in the image directly affects the Peak Signal to Noise Ratio and the statistical properties of the image. These embedding algorithms are applicable mainly to lossless image compression schemes like TIFF images. For lossy compression schemes like JPEG, some of the message bits get lost during the compression step.

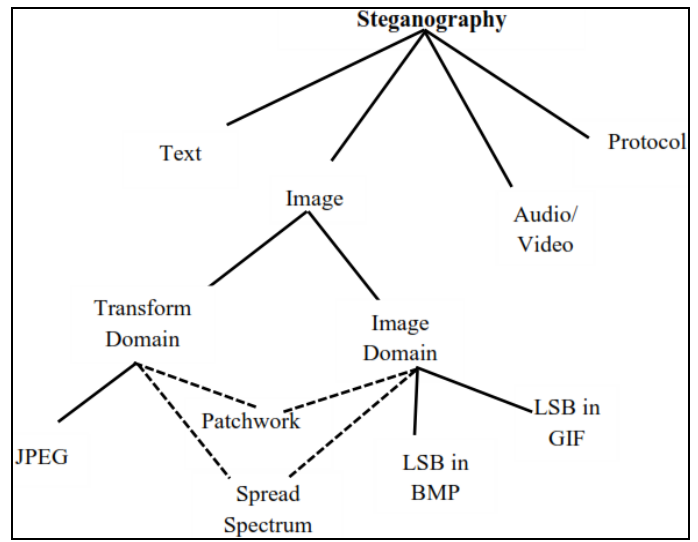

Fig. 2. Categories of image Steganography

Fig. 2 presents the detail flow of the image steganography. The Least Significant Bit (LSB) Replacement technique is the most common algorithm belonging to this class of techniques. As suggested by the name itself, the least significant bit of the binary representation of the pixel gray levels is replaced to represent the message bit. The replacement of the LSB causes an addition of a noise in the pixels of the image.

This kind of embedding also leads to an asymmetry and a grouping in the pixel gray values (0, 1) ;( 2, 3). . . (254,255). To overcome this undesirable asymmetry, the decision of changing the least significant bit is randomized i.e. if the message bit does not match the pixel bit, then pixel bit is either increased or decreased by 1. This technique is popularly known as LSB Matching [3].

D. *Transform Domain Steganography*

While spatial domain steganography uses the image properties directly, the transform domain steganography encodes the message bits in the transform domain coefficients of the image. Data embedding performed in the transform domain is widely used for robust watermarking. Similar techniques can also realize large capacity embedding for steganography. Discrete Cosine Transform (DCT), Discrete Wavelet Transform (DWT), and Discrete Fourier Transform (DFT) are the most popular candidate transforms for this type of steganography.

Being embedded in the transform domain, the hidden data resides in more robust areas, spread across the entire image thus providing better resistance against signal processing. For example, we can perform a block DCT and choose one or more components in each block to form a new data group depending on payload and robustness requirements. The image is pseudo randomly scrambled and undergoes a second-layer transformation. Using various transform domain schemes, modification is then carried out on the double transform domain coefficients previously derived. These techniques have high embedding and extraction complexity. Because of the robustness properties of transform domain embedding, these techniques are generally more applicable to the "Watermarking" aspect of data hiding [3].

### III. EXISTING METHODS

Steganography may sound new to many people but it is not a new science. Methods used may be different but it has been practiced since ancient periods for secret communication.

Familiar steganography techniques include invisible inks, the use of carrier pigeons, and the microdot. Along with the development of digital communication, newer and more effective steganography techniques came into existence. Moving with the pace of digitization, more of today's communication occurs electronically. As a result, there have been advancements in steganographic techniques through the use of digital multimedia signals such as audio, video or image as cover signals. In traitor-tracing schemes, the original cover signal is needed to reveal the hidden information and such schemes are called cover escrow. In this scenario, the assignee's identification is embedded in the copies of the cover signal, thereby producing modified cover signal and disseminated. In case of acquisition of the illegal copies of the signal, subtraction of the original cover data from the modified signal results in the source of copy plus the offender's identity. However, in many applications it is not practical to require the possession of the unaltered cover signal to extract the hidden information. Another variant, the blind schemes allow direct extraction of the embedded data from the modified signal without knowledge of the original cover. Blind strategies are predominant among steganography of the present day [4].

A block diagram of a Steganographic system is depicted in Fig. 3. An original text is encrypted and then embedded in a digital image by the stegosystem encoder using a stego-key. The resulting stego image is transmitted over a channel to the receiver, where it is processed by the stegosystem decoder using the same key. During transmission, the stego image can be monitored by eavesdroppers/viewers who can only observe the image itself without even noticing the fact that a secret message lies within it.

Since steganography leaves behind detectable traces (i.e., distortion) in the stego object and modifies the statistical properties, steganalysis is used to detect the presence of distorted statistical properties. The statistical attacks can further be classified as: Targeted Attacks and Blind Attacks. Steganalysis is a relatively new research discipline with few articles appearing before the late-1990s. Basically steganalysis is the process of detecting steganography by looking at variances between bit patterns and unusually large file sizes. It is the art of discovering and rendering useless covert messages as mentioned in a paper by Satenik et al. [5]. The goal of steganalysis is to identify suspected information streams, determine whether or not they have hidden messages encoded into them, and, if possible, recover the hidden information.

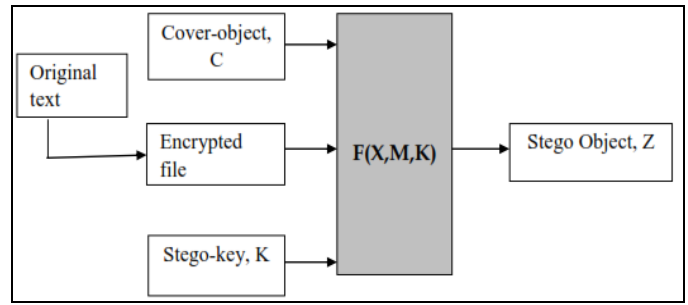

Fig. 3. Steganography combined with encryption

The steganalysis techniques focus on detecting the presence/absence of a secret message in observed message. To our knowledge, there seems to have been no attempt in extracting the secret message. Diop et al. in their paper have mentioned that the extraction of the secret message could be a harder problem than mere detection [6]. Therefore, based on the ultimate outcome of the effort, steganalysis has been classified into two categories:

Passive steganalysis: Detect the presence or absence of a secret message in an observed message and

Active steganalysis: Extract a (possibly approximate) version of the secret message from a stego message.

So as to assure reliability and prevent the loss of original information hidden in the digital data, it is necessary to introduce an appropriate framework that can encrypt and hide the contents. Some works have been carried out subsequently on the related filed. Sutaone et al. in [7] designed steganography system for encoding and decoding a secret file embedded into an image file using random LSB insertion method in which the secret data were spread out among the image data in a seemingly random manner. To achieve this, they have used a secret key so as to generate the pseudorandom numbers, which would identify where, and in what order the hidden message was laid out. Their work just incorporated some cryptography in that diffusion was applied to the secret message.

Neil et at. in [8] have discussed three popular methods for digital message concealment which were LSB insertion, masking and filtering and algorithmic transformation. Kevin et al. in their paper [9] analyzed seven different image steganography methods. These methods were Stego1bit, Stego2bits, Stego3bits, Stego4bits, Stego ColourCycle, StegoPRNG, and StegoFridrich.

Several digital data hiding techniques for images such as substitution systems, transform domain techniques, statistical steganography, hiding in two -color images, distortion and cover generation were explored, analyzed, attacked and finally counterpart was provided by Neil F. Johnson et al. in [10]. Wu et al. presented an adaptive steganographic scheme based on pixel-value differencing method [11]. Accordingly, the hiding capacity of each pixel could be different. Since the degree of distortion tolerance of an edge area was naturally higher than that of smooth area, instead of smooth areas they chose edge areas or pixels to hold more secret data.

Similarly, Chun-Shien Lu in [12] presented several techniques for steganography, watermarking, fingerprinting, signature based image authentication for digital image and audio files. Alwan et al. introduced a novel approach of image embedding on 8-bit images [13]. The proposed method embedded three images and text in one image using edge-pixels. Out of the several, only LSB insertion method was used in the implementation of the techniques proposed in the paper by Juneja et al. [14]. Their work focused on a new image-based triple-A algorithm. They used the same principle of LSB, where the secret message bit was hidden in the least significant bits of the pixels, with more randomization in selection of the number of bits used and the color channels that were used. The randomization was expected to increase the security of the system and to increase the capacity. The technique can be applied to RGB images where each pixel is represented by three bytes to indicate the intensity of red, green, and blue in that pixel.

However in this paper, we have used AES for encryption of the original message, and SDSA to hide .txt file inside the image.

## IV. SPATIAL DESYNCHRONIZATION

To break the JPEG steganographic algorithms, the calibration attack during eavesdropping has been one of the most successful attacks in recent past. The successful prediction of cover image statistics from a stego image leads to the success of the eavesdropping. It is necessary to prevent the eavesdropper from successfully predicting the cover image statistics in order to resist the calibration attack. The separation of the embedding domain from the staganalytic domain is used to prevent cover image prediction from the stego image. In other words, if the embedding domain is kept secret from the eavesdropper then it is not possible to mount calibration attack by predicting the cover image statistics [15].

The spatial de-synchronization operation randomizes the embedding domain so as to hide it from the eavesdropper. Here spatial desynchronization implies the embedding grid is not synchronized with the JPEG compression grid of the stego image. The spatial shifting (desynchronization) adds up a noise (sometimes called desynchronization noise) to the stego image and the noise masks the steganographic noise in such a way that the detection of steganographic embedding becomes difficult for the eavesdropper [16]. Thus spatial shifting operation resists the attempted calibration attacks.

### A. Spatial Block Desynchronization

An image is divided into non-overlapping blocks of size 8×8 and then the information contained in these blocks is then compressed by taking the 2D Discrete Cosine Transform of the block followed by quantization step which are then used for embedding data bits. A slight alteration of this spatial block arrangement can desynchronize the whole image, thus termed as Spatial Block Desynchronization. For example, 8×8 non overlapping blocks for embedding can be taken from a subimage of the original cover image or we can say the block arrangement is slightly shifted from standard JPEG compression block arrangement. A formal description of spatial block desynchronization is given below.

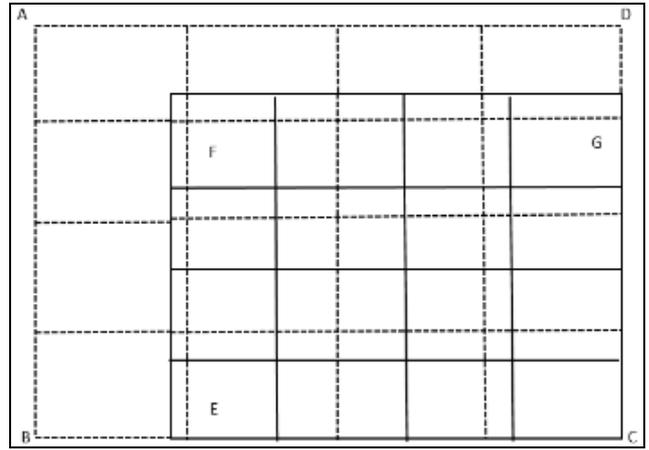

Fig. 4. Block Diagram of Spatial Block Desynchronization

Let I be a gray scale image of size (N×N). A sub image of I can be obtained by removing u rows from the top and v columns from left. Let us denote the cropped image by $I_c$ u, v with size (N-u) × (N-v) and the cropped out portion of image $I_o$ u, v can be represented as:

$I_o$ u, v = I – $I_c$ u, v

The block size should not necessarily be 8×8, i.e. Spatial domain desynchronization presents the choice of using blocks of sizes m×n where m ≠ 8 and n ≠ 8. In such a case, the quantization matrix Q has to be changed accordingly to size m×n at the time of data embedding. Randomization can further strengthen the desynchronization process. In this case, the removal of rows and columns and also the sizes of the blocks can be chosen randomly using a shared secret key and matrix Q between the two communicating parties. In the steganalysis process, the image statistics are derived using blocks of sizes 8×8. Choosing the block size other than 8× 8, the steganalyst is not able to capture effectively the modifications made during the embedding process. It is difficult to track the portions of the image containing the embedded information due to randomized hiding, even if it is known that embedding has been done using blocks of different sizes. Any JPEG steganographic scheme can be employed for embedding, once the quantized DCT coefficients have been obtained.

## V. PROCEDURE AND RESULTS

The Spatially Desynchronized Steganography Algorithm (SDSA) has been used to embed data in a spatially desynchronized version of the cover image so that the cover image statistics cannot be easily recovered from the stego image. Fig. 5 shows the spatial desynchronization scheme. The cropped version of the image $I_c$ u, v is used for Steganographic embedding using any DCT domain scheme. After embedding, this embedded portion of the image is stitched with $I_o$ u, v to obtain the stego image $I_s$. The JPEG compressed version of $I_s$ is communicated as the stego image.

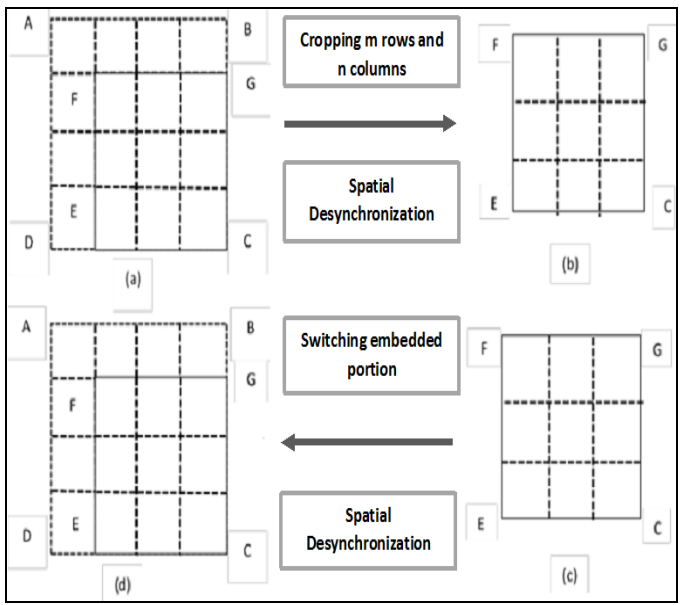

Fig. 5.  Spatial Desynchronization

Below is a stepwise description of the algorithm [3]:

Input: Cover Image I

Input Parameters: Rows and Columns to be cropped (u, v), Block size (m×n), Quantization Matrix (Q)

Output: Stego Image $I_s$

Begin

1. Partition the cover image I into $I_c$ u, v and $I_o$ u, v by cropping u topmost rows and v leftmost columns.

2. Perform m×n non-overlapping block partitioning on $I_c$ u,v. Let us denote this set of blocks by $P_{I\,u,\,v}$ (m×n).

3. Choose a set of blocks from $P_{Ic\,u,\,v}$ (m × n) (using a key shared by both ends) and perform the embedding in each of the selected blocks using any standard DCT based steganographic scheme. The quantization matrix Q which is a shared secret is used for obtaining the quantized coefficients.

4. Apply dequantization and Inverse Discrete Cosine Transform (IDCT) to the set of blocks used for embedding in Step 3.

5. Join $I_o$ u,v with the resulting image obtained at Step 4. This combined image is the output stego image $I_s$ which is compressed using JPEG compression and communicated as the stego image.

End.

For the encryption of the data, Advanced Encryption Standard (AES) was used. Overall encryption and decryption via AES has been shown in Fig. 6. AES, developed by two Belgian cryptographers, Joan Daemen and Vincent Rijmen, is a symmetric block cipher based on Rijndael cipher that is intended to replace DES as the approved standard for a wide range of applications.

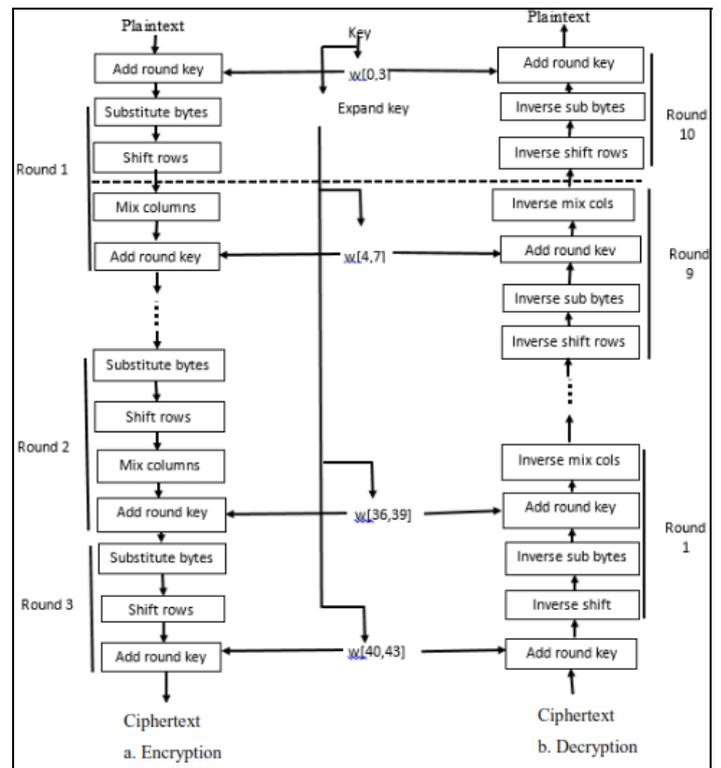

Fig. 6.  AES

For AES, NIST selected three members of the Rijndael family, each with a block size of 128 bits, but three different key lengths: 128, 192 and 256 bits. The key size used for an AES cipher specifies the number of repetitions of transformation rounds that convert the input, called the plaintext, into the final output, called the ciphertext. The number of cycles of repetition are as follows:

1. 10 cycles of repetition for 128-bit keys.
2. 12 cycles of repetition for 192-bit keys.
3. 14 cycles of repetition for 256-bit keys.

Four different stages are used, one of permutation and three of substitution:

1. Substitute bytes: Uses an S-box to perform a byte-by-byte substitution of the block
2. ShiftRows: A simple permutation
3. MixColumns: A substitution that makes use of arithmetic over $GF(2^8)$
4. AddRoundKey: A simple bitwise XOR of the current block with a portion of the expanded key

Unit testing has been carried out on each module before integrating them. Since Visual Studio was used, simple syntactical mistakes were immediately shown so that the correction was of much ease. The research work resulted in the development of a desktop application that has specially been designed for police department where the department administrator has the full authority to all the services. After the login verification for admin, the required information can be written, the secret information is then encrypted thereby forming ciphertext which can be hidden in the jpeg image. The webcam is used to capture the photograph for records and the image containing the hidden ciphertext are sent via email to the specified departments. Then the hidden secret messages can be extracted from the digital images.

Basically, the size of the stego-image depends on the length of the text. For example,

Size of original image: 659 KB
For long length text, size of stego-image: 666 KB
For short length text, size of stego-image: 660 KB

The following Fig. 7, Fig. 8, Fig. 9 and Fig. 10, illustrate the original message, encrypted text, original image and the image with hidden message respectively. The size of original text message was 25 bytes, the size of the encrypted text message was then 88 bytes, the size of original image was 194,971 bytes and the size of image with hidden ciphertext became 1,688,201 bytes. Fig. 11 presents the snapshot of our deveoped application that encrypts, hides and retrives the message and maintains database of the required records.

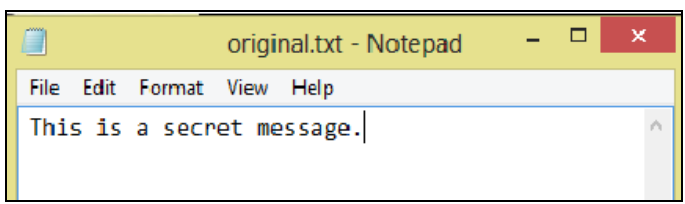

Fig. 7.   Original Text Message

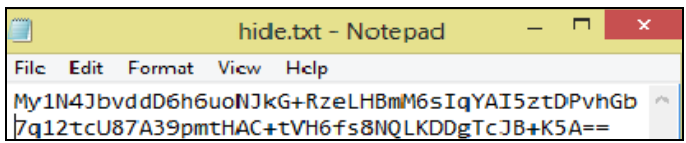

Fig. 8.   Encrypted Text

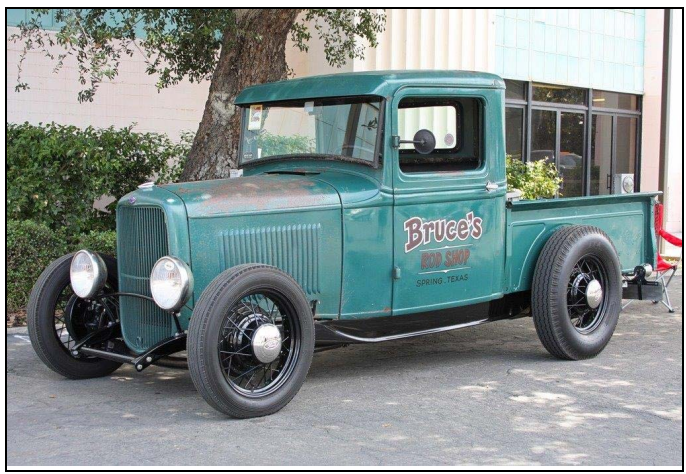

Fig. 9.   Original image

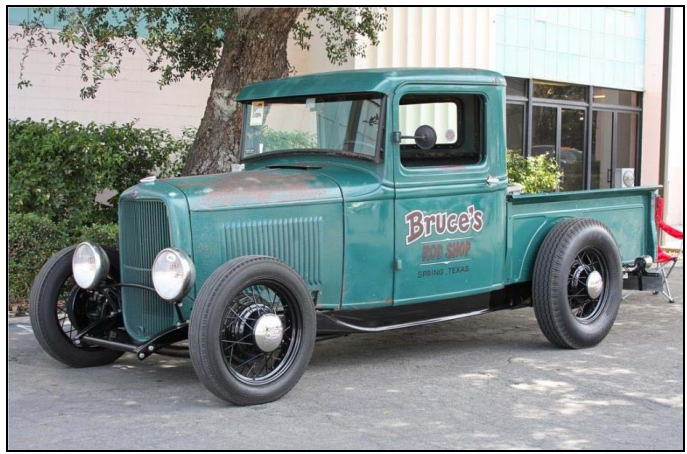

Fig. 10. Image with hidden ciphertext

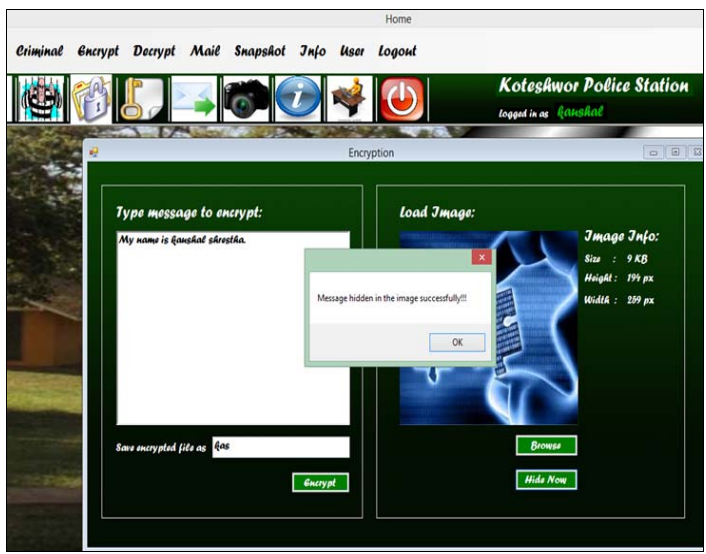

Fig. 11. Image with hidden ciphertext of the apps

VI.   CONCLUSION

The secure communication of information is of much essence in today's world. Day per day eavesdropping are going on during the confidential information transmission; may it be the banking details of a bank, the call taping of people for blackmailing, the information leakage through unsecure communication media etc. Thus, concerning about these aspects, Image Steganography is the application for the organizations such as police department to hide confidential police details like strictly confidential criminal details, secret missions, police strategy etc. inside an image file. Also this technique can be widely applicable to deliver required confidential matters even through social networking sites where people will not even think that the image could hold something else. The message is highly secure since it has double-layer protection: encryption and steganography, and has wide range of applications in various sectors.